\documentclass{article}
\usepackage{spconf,amsmath,graphicx}
\usepackage{subcaption}
\usepackage{url}
\usepackage{cite}
\usepackage{amsmath,amssymb,amsfonts}
\usepackage{multirow}
\usepackage{siunitx}
\usepackage{algorithmic}
\usepackage{graphicx}
\usepackage{textcomp}
\usepackage{xcolor}
\usepackage{amsmath}
\usepackage{hyperref}
\usepackage{comment}
\usepackage{url}

\usepackage{enumitem}
\setlist{nosep, leftmargin=14pt}

\usepackage{mwe} 

\title{FOD-Swin-Net: angular super resolution of fiber orientation distribution using a transformer-based deep model}
\name{Mateus Oliveira da Silva, Caio Pinheiro Santana, Diedre Santos do Carmo, Letícia Rittner}
\address{School of Electrical and Computer Engineering, Universidade Estadual de Campinas, Campinas, Brazil}

\usepackage{tikz}

\newcommand\submittedtext{%
  \footnotesize © 2024 IEEE. Personal use of this material is permitted. Permission from IEEE must be obtained for all other uses, in any current or future media, including reprinting/republishing this material for advertising or promotional purposes, creating new collective works, for resale or redistribution to servers or lists, or reuse of any copyrighted component of this work in other works.
}

\newcommand\submittednotice{%
  \begin{tikzpicture}[remember picture,overlay]
    \node[anchor=south,yshift=10pt] at (current page.south) {\fbox{\parbox{\dimexpr\textwidth-\fboxsep-\fboxrule\relax}{\submittedtext}}};
  \end{tikzpicture}%
}

\begin{document}
%
\maketitle
\begin{abstract}

\submittednotice

Identifying and characterizing brain fiber bundles can help to understand many diseases and conditions. An important step in this process is the estimation of fiber orientations using Diffusion-Weighted Magnetic Resonance Imaging (DW-MRI). However, obtaining robust orientation estimates demands high-resolution data, leading to lengthy acquisitions that are not always clinically available. In this work, we explore the use of automated angular super resolution from faster acquisitions to overcome this challenge. Using the publicly available Human Connectome Project (HCP) DW-MRI data, we trained a transformer-based deep learning architecture to achieve angular super resolution in fiber orientation distribution (FOD). Our patch-based methodology, FOD-Swin-Net, is able to bring a single-shell reconstruction driven from 32 directions to be comparable to a multi-shell 288 direction FOD reconstruction, greatly reducing the number of required directions on initial acquisition. Evaluations of the reconstructed FOD with Angular Correlation Coefficient and qualitative visualizations reveal superior performance than the state-of-the-art in HCP testing data. Open source code for reproducibility is available at \url{https://github.com/MICLab-Unicamp/FOD-Swin-Net}.
\end{abstract}
\section{Introduction}
\label{sec:intro}

Diffusion-Weighted Magnetic Resonance Imaging (DW-MRI) can reveal details of white matter (WM) microstructure by measuring the random movement of water molecules in the tissue. In particular, estimating the orientation of the main fiber bundles of the brain from DWI (Diffusion-weighted images) has been used for various clinical and research applications. Through these estimates, it is possible to carry out pre-operative planning for gliomas, analyzing fiber bundles using tractography~\cite{preoperative_glioma}. Moreover, DWI is useful for assessing neuro-degenerative diseases, such as amyotrophic lateral sclerosis and Huntington's disease, allowing for the verification of the integrity of fiber pathways and understanding disease progression~\cite{neuro_degenerative_overview}. The technique is also employed in evaluating strokes, to determine the extent of damage to the fiber pathways~\cite{stroke_chronic_fibers}.

To obtain estimates of fiber bundle orientation, some form of modeling needs to be applied to the DWI data. Despite its prevalence in literature and clinical practice, the diffusion tensor model~\cite{basser_mr_1994} depicts only one principal diffusion direction. Therefore, it is not adequate to represent brain regions with more than one fiber orientation, leading to errors in fiber tractability and the identification of connections that do not match reality~\cite{CSD_multi_tissue}. Several higher-order models have been proposed to overcome this limitation. Multi-tensor fitting methods define a discrete set of fiber bundles, while Spherical Deconvolution (SD) methods define a continuous fiber orientation distribution (FOD). Currently, the SD models are among the state-of-the-art for tractography applications due to their ability to handle complex fiber configurations efficiently, especially in High Angular Resolution Diffusion Imaging (HARDI) acquisitions~\cite{tournier_diffusion_2019,dell2019modelling}. 

\begin{figure*}[t]
\centering
\includegraphics[width=0.9\textwidth]{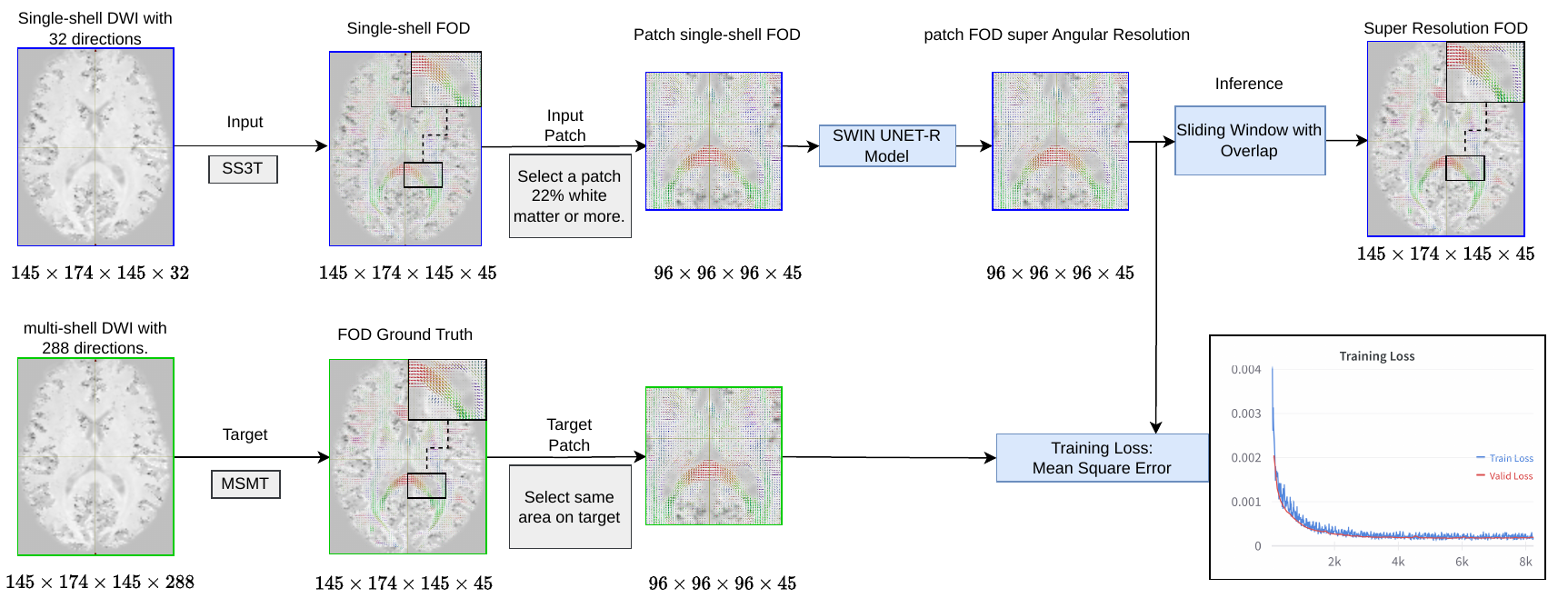}

\caption{
Outline of the method. The downsampled DWI is turned into single-shell FOD using SS3T. 3D patches with significant brain white/gray matter presence are selected to train a Swin UNETR architecture. Targets for FOD reconstruction learned with MSE loss are derived from multi-shell 288 directions data. A whole volume is constructed through sliding windows.}
\label{fig:arquitetura}
\end{figure*}

Among SD methods, the Constrained SD (CSD)~\cite{tournier2007robust} uses the spherical harmonic (SH) basis to represent the FOD, applying a non-negativity constraint that improves noise robustness. Single-Shell Three-Tissue CSD (SS3T CSD)~\cite{single_shell_three_tissue} extends CSD to model three tissue types, making it well-suited for single-shell acquisitions and WM regions with partial volume effect. Regarding its usage of single-shell acquisitions, recall that the b-value is a factor that reflects the strength and timing of the gradients used to generate DWIs. Single-shell acquisitions, which frequently use b-values around 1000 $s/mm^2$, cannot adequately characterize the relative contribution of the SH coefficients from lower to higher harmonic degrees. This makes acquisitions with multiple b-values, known as multi-shell acquisitions, more relevant. Multi-shell acquisition provides better angular contrast~\cite{b_effect_determination,b_effects_2_pulse_duration} and can be incorporated into the CSD approach through the Multi-Shell Multi-Tissue CSD (MSMT CSD) method~\cite{CSD_multi_tissue}.

Although multi-shell HARDI acquisitions are desirable, they are not always available in clinical settings. Recent works have proposed deep learning (DL) methods to overcome this limitation. In~\cite{bartlett2023recovering}, FODs are reconstructed from multi-shell DWI data with a reduced number of directions using a model-based architecture. More specifically, consistency blocks exploiting prior knowledge from the CSD model are combined with convolutional deep regularization blocks to obtain solutions consistent with the input data. In FOD-Net~\cite{zeng2022fod}, voxel-wise Convolutional Neural Networks (CNN) are used to enhance the angular resolution of FOD images computed from single-shell Low Angular Resolution Diffusion Imaging (LARDI) data.

This paper presents a method that employs the SS3T algorithm for the initial reconstruction of the FOD and feeds it to a deep neural network with transformer attention mechanisms for generating FOD super resolution. The proposed FOD-Swin-Net, based on the Swin UNETR~\cite{swin_unet_r}, learns to refine the spherical harmonic coefficients from 3D patches extracted from relevant regions. The model is able to effectively bring FOD constructed from single-shell 32-direction DWI to have comparable quality to FOD constructed from multi-shell 288 directions, achieving superior performance when evaluated against FOD-Net, a state-of-the-art super resolution method. 

\section{Dataset}
The data used for training and evaluating the proposed model comes from the Human Connectome Project (HCP)~\cite{van2013wu}, a consortium studying brain connectivity in healthy adults that made data publicly available, including T1w and DW images. The DWI was acquired using b-values = 1000/2000/3000 $s/mm^2$ with 90 non-collinear gradient directions for each b-value and 18 volumes with b-value = 0 $s/mm^2$ ($b_0$). In total, 131 subjects from the HCP were selected.

It is important to mention that the original angular directions of the HCP data were down-sampled for serving as input to the implemented methods. While each subject originally had 288 diffusion directions (multi-shell), we reduced this number to 32 + $b_{0}$ directions (single-shell) based on the directionality reduction methodology~\cite{design_aquisition_HCP}, i.e., following the original acquisition guidelines from HCP (Table~\ref{table:info_down_sampling}).

\begin{table}[ht]
\centering
\begin{tabular}{|c|c|c|c|}
\hline
\textbf{b}    & \textbf{N} & \textbf{LARDI data}            & \textbf{HARDI data} \\ \hline
1000 & 90     & select 32 directions  & select all \\ \cline{1-4} 
2000 & 90     & ignore                & select all \\ \cline{1-4} 
3000 & 90     & ignore                & select all \\ \cline{1-4} 
0    & 18     & select a $b_0$ volume & select all \\ \hline
\end{tabular}
\caption{Details on how the LARDI (input) and HARDI (target) data were selected from each of the 131 subjects to compose the final dataset. N represents the number of directions or $b_0$ acquisitions.  
}
\label{table:info_down_sampling}
\end{table}

IDs for the selected scans are available in our GitHub repository. The FSL decomposition provided by MRtrix~\cite{tournier_mrtrix3} was used to compute the three tissue masks: WM, Subcortical Gray Matter (SGM), and Cortical Gray Matter (CGM).  

\section{Proposed model}

The downsampled input, a single-shell data with dimensions \(145 \times 174 \times 145\) and 32 directions along with an additional $b_0$ volume, is fit using the SS3T  algorithm~\cite{single_shell_three_tissue}, producing a FOD with dimensions \(145 \times 174 \times 145 \times 45\), being 45 the coefficients of the FOD. Patches of size \(96 \times 96 \times 96 \times 45\) are then randomly extracted, assuring that they contain at least 20\% of SGM, CGM, or WM. The Swin UNETR~\cite{swin_unet_r} fed with the described input patches produces a patch with equivalent dimensions, \(96 \times 96 \times 96 \times 45\), which is compared and optimized through Mean Square error with a multi-shell MSMT CSD derived target. The full resolution output during evaluation is reconstructed through a sliding window with an overlap of \(25\%\) (Fig.~\ref{fig:arquitetura}).

Swin UNETR~\cite{swin_unet_r} is a type of vision transformer network that adopts a hierarchical approach to partition the image into sub-patches. As the network deepens, these sub-patches are merged, allowing the model to efficiently capture both local and global features of the image. This progressive merging mechanism helps maintain a balance between local and global information, and we hypothesize that this balance is crucial for the proper angular super resolution of the input patch.
\vspace{6pt}
\newline
\textit{Model training}

From the 131 subjects selected from the HCP dataset, 71 subjects were selected for training, 10 for validation, and 50 for test. The size of the validation and test sets were chosen to be the same as the FOD-Net, for comparison purposes. The size of the training set, however, is larger than the original article of FOD-Net. We chose to expand the training dataset because transformer-type models typically require more data to achieve better convergence. 

The network was trained using the ADAM optimizer with a learning rate of $0.0005$ and batch size specified as 2. Training was carried out for 80 epochs until convergence, using the criterion of the lowest validation error. When it reached a stationary region, the training was halted.
\vspace{6pt}
\newline
\textit{Evaluation}

In order to assess the resulting FODs and enable a direction comparison with other methods (SS3T and FOD-Net), the spherical harmonic coefficients were evaluated using the Angular Correlation Coefficient (ACC)~\cite{acc_anderson}, following the formulation used by FOD-Net~\cite{zeng2022fod}. The tissue masks were used to perform evaluations only in fibrous tissue, avoid measuring errors in noisy regions, and assess the impact of voxels with partial volume on the results:

\begin{itemize}
    \item Voxels containing at least \( 70\% \) of WM.
    \item Voxels containing at least \( 30\% \) of WM and \( 30\% \) of CGM.
    \item Voxels containing at least \( 30\% \) of WM and \( 30\% \) of SGM.
\end{itemize}

\section{Results}

\begin{figure}[h!]
 \centering
  \begin{tabular}[h]{c}
    \includegraphics[width=0.9\linewidth]{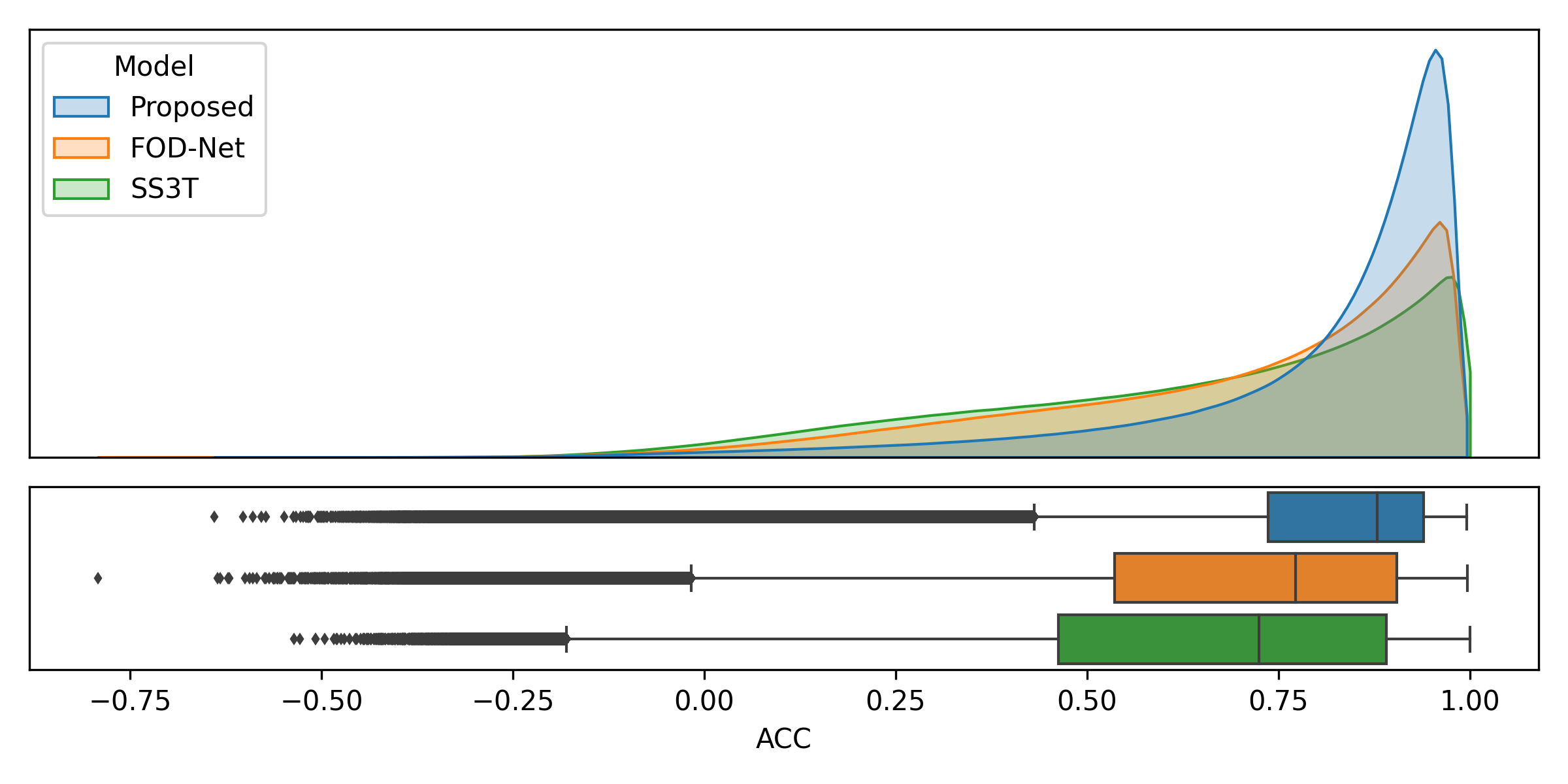} \\
    \small (a) WM ACC distribution.
  \end{tabular}
  \begin{tabular}[h]{c}
    \includegraphics[width=0.9\linewidth]{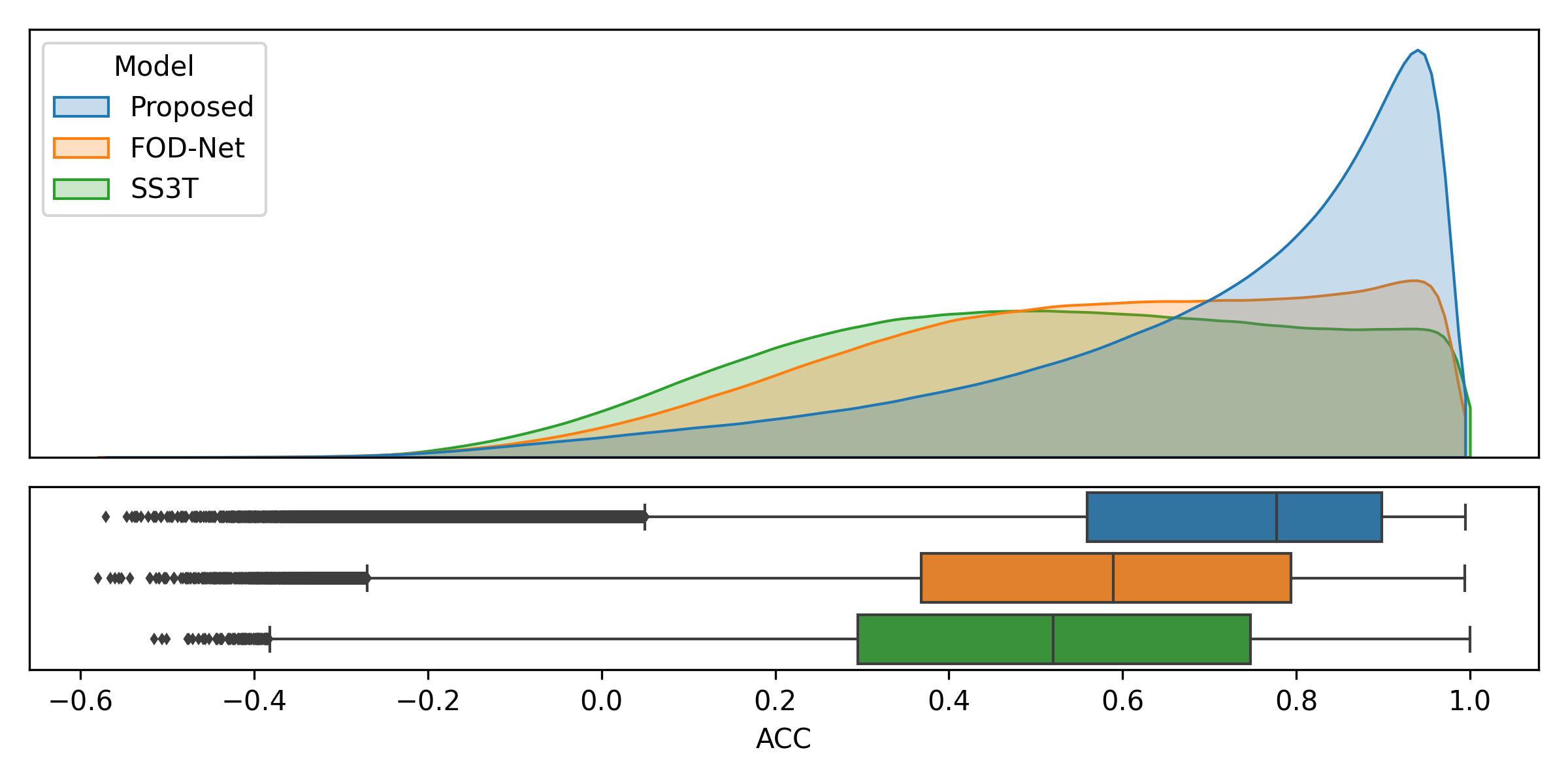} \\
    \small (b) WM and CGM ACC.
  \end{tabular}
  \centering
  \begin{tabular}[h]{c}
    \includegraphics[width=0.9\linewidth]{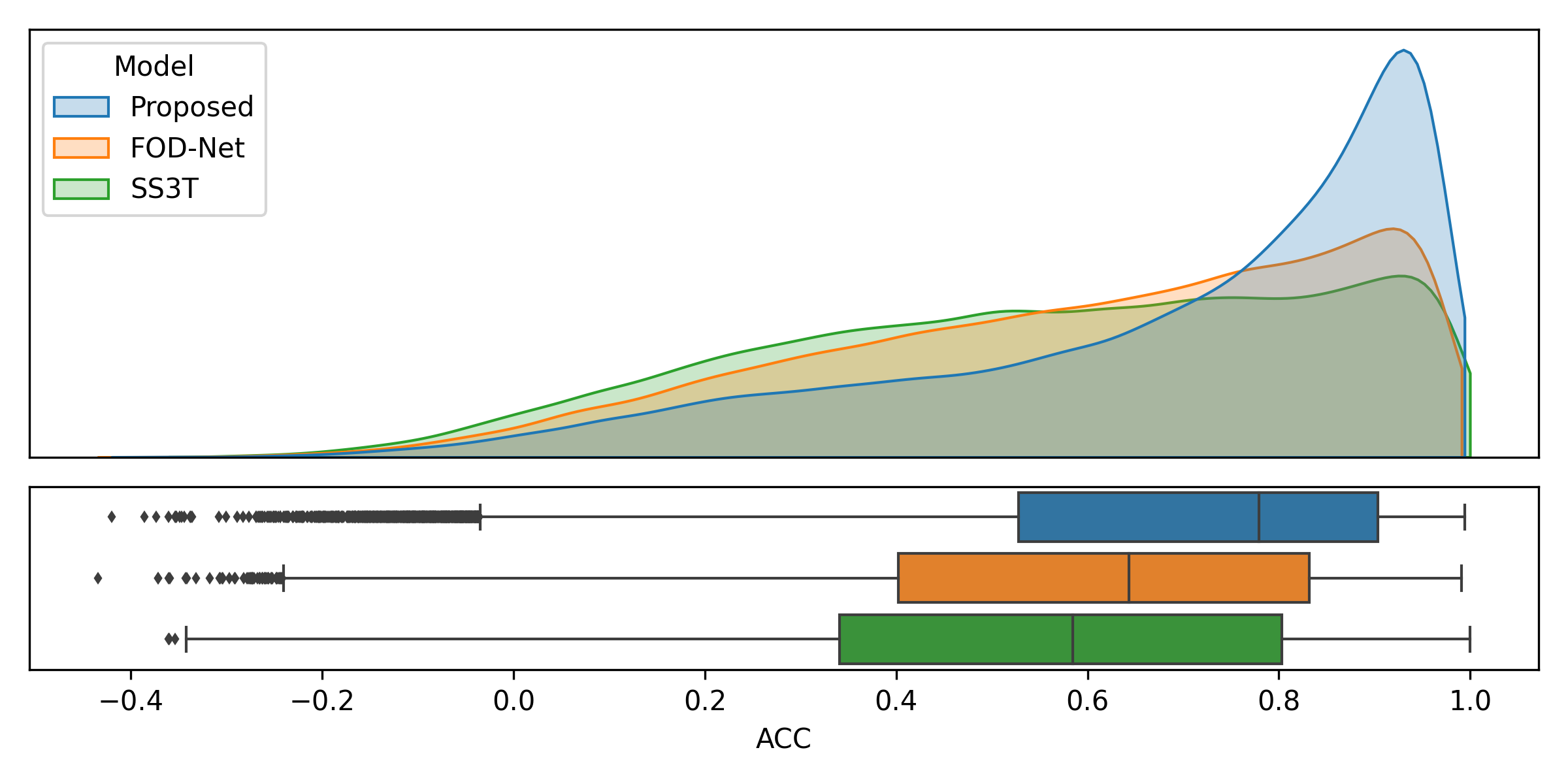} \\
    \small(c) WM and SGM ACC.
  \end{tabular} 
   
  \caption{Half-violin distributions and boxplots for each method's ACC in the test set, in different regions: WM, WM with CGM, and WM with SGM.}
  \label{fig:acc_violin_boxplot}
\end{figure}

 In this study both FOD-Net and FOD-Swin-Net were trained with 71 subjects, to ensure a fair comparison. Comparisons of FOD coefficient ACC obtained across three different tissue regions demonstrate the superior performance of FOD-Swin-Net (Fig.~\ref{fig:acc_violin_boxplot} and Table~\ref{metric:table_acc}). Distribution of ACC across all subjects in the test set shows superior mean performance with less variance of FOD-Swin-Net over FOD-Net and the baseline SS3T reconstruction. Regarding computational time, the proposed patch-based approach is more efficient for inference. Our reproduction of FOD-Net takes around 60 seconds to infer all super resolution voxels of a single subject, while FOD-Swin-Net performs the full sliding window in 1 second, in the same hardware (Intel I9 13900KF and Nvidia 4090 GPU). 

\begin{table}[h!]
\centering
\resizebox{\columnwidth}{!}{
\begin{tabular}{cccccccc}
\hline
\textbf{Model} & \textbf{Tissue} & \textbf{Min} & \textbf{Max} & \textbf{Mean} & \textbf{STD} & \textbf{LQ} & \textbf{UQ} \\
 & \textbf{Type} &  & & & & \textbf{} & \textbf{} \\ \hline
SS3T & WM & -0.54 & 1.00 & 0.65 & 0.28  & 0.46 & 0.89 \\
CSD & WM/CGM & -0.52 & 1.00 & 0.51 & 0.29 & 0.30 & 0.75 \\
& WM/SGM & -0.36 & 1.00 & 0.56 & 0.29 & 0.34 & 0.80 \\ \hline
FOD- & WM & -0.79 & 1.00 & 0.69 & 0.26 & 0.54 & 0.90 \\
Net & WM/CGM & -0.58 & 0.99 & 0.56 & 0.27 & 0.37 & 0.79 \\
& WM/SGM & -0.43 & 0.99 & 0.60 & 0.27 & 0.40 & 0.83\\ \hline
FOD- & WM & -0.64 & 1.00 & \textbf{0.80} & 0.21 & 0.74 & 0.94 \\
Swin- & WM/CGM & -0.57 & 0.99 & \textbf{0.69} & 0.26 & 0.56 & 0.90 \\
Net & WM/SGM & -0.42 & 0.99 & \textbf{0.69} & 0.27 & 0.53 & 0.90\\ \hline
\end{tabular}
}
\caption{Statistics of ACC in three different tissue types from SS3T and FOD-Net compared to our method. LQ and UQ stand for lower and upper quartiles, respectively.}
\label{metric:table_acc}
\end{table}

\begin{figure*}[ht]
    \centering
    \begin{subfigure}{0.145\textwidth}
        \includegraphics[width=\textwidth]{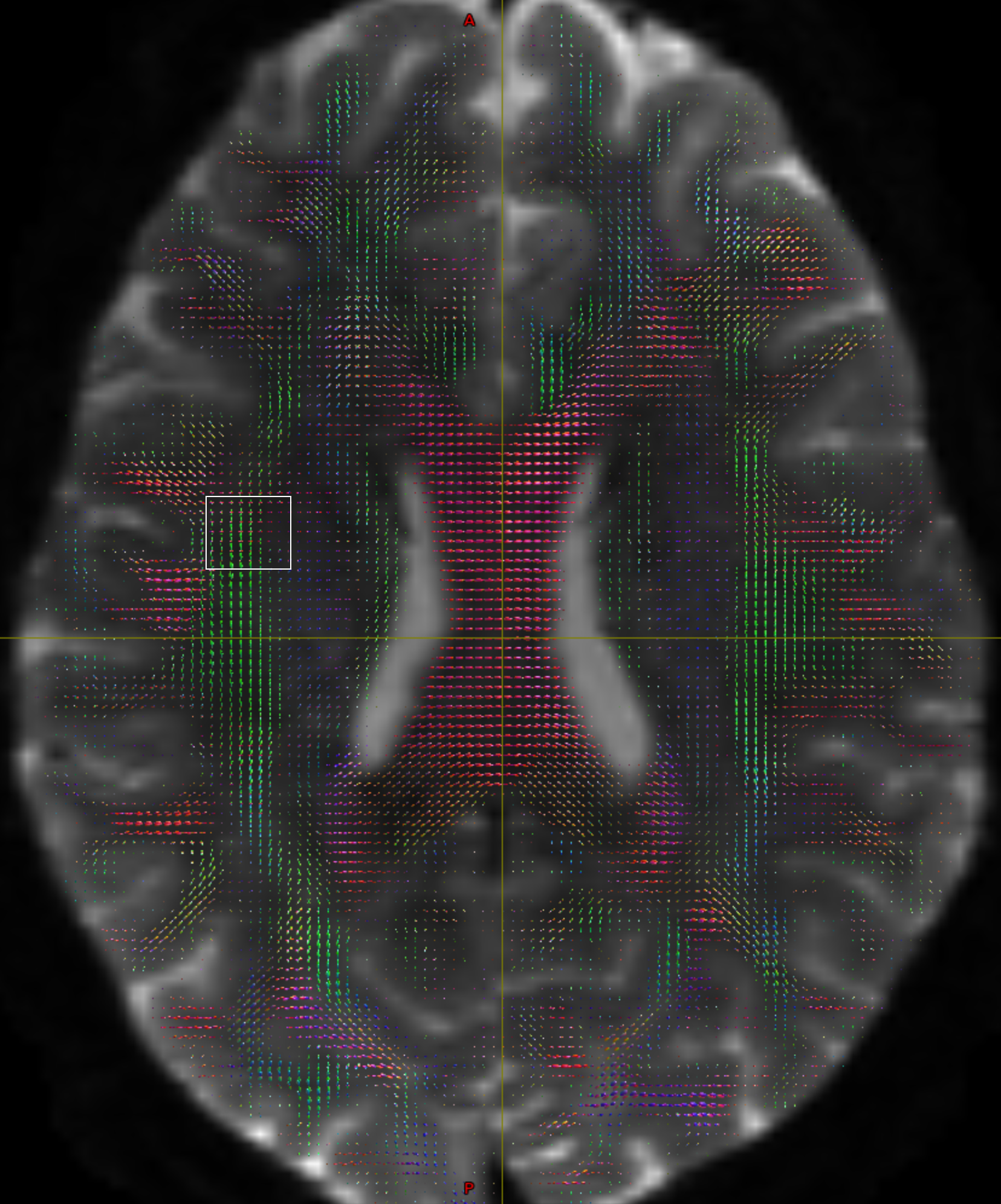}
        \caption{Selected region}
    \end{subfigure}
    \hspace{0.01mm}
    \begin{subfigure}{0.17\textwidth}
        \includegraphics[width=\textwidth, trim=0 0 480 50, clip]{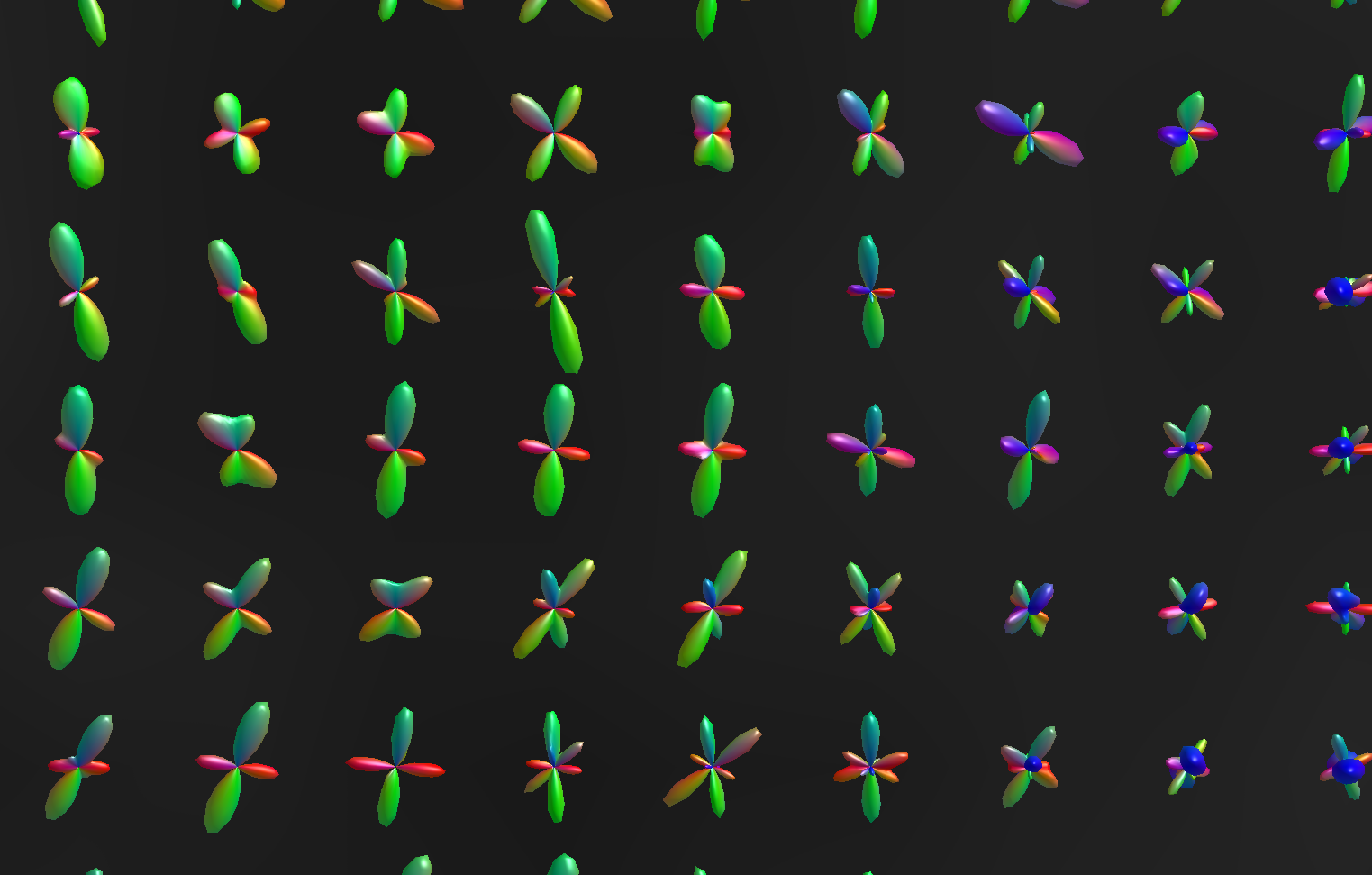}
        \caption{SS3T}
    \end{subfigure}
    \hspace{0.01mm}
    \begin{subfigure}{0.17\textwidth}
        \includegraphics[width=\textwidth, trim=0 0 480 50, clip]{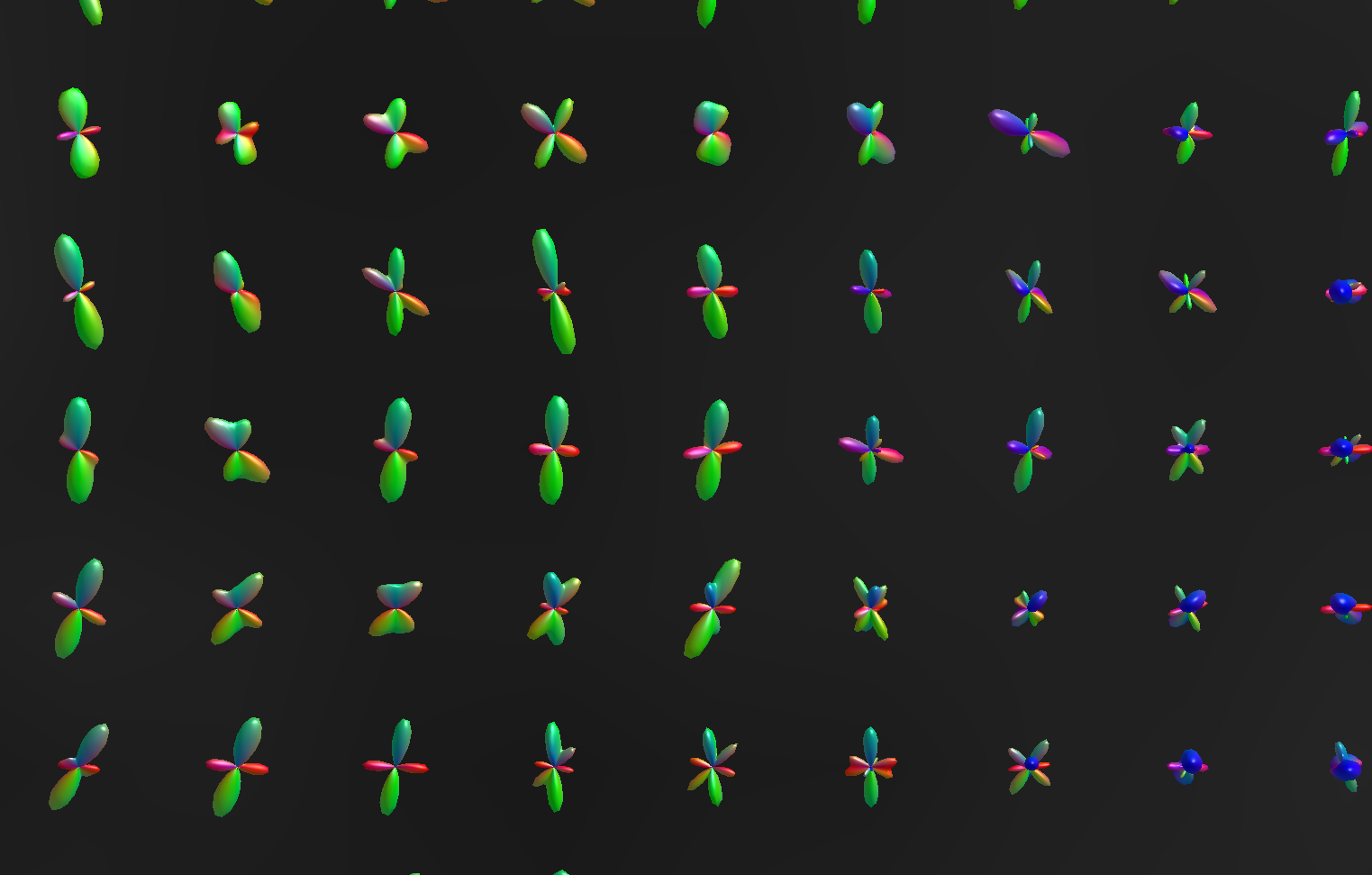}
        \caption{FOD-Net}
    \end{subfigure}
    \hspace{0.01mm}
    \begin{subfigure}{0.17\textwidth}
        \includegraphics[width=\textwidth, trim=0 0 480 50, clip]{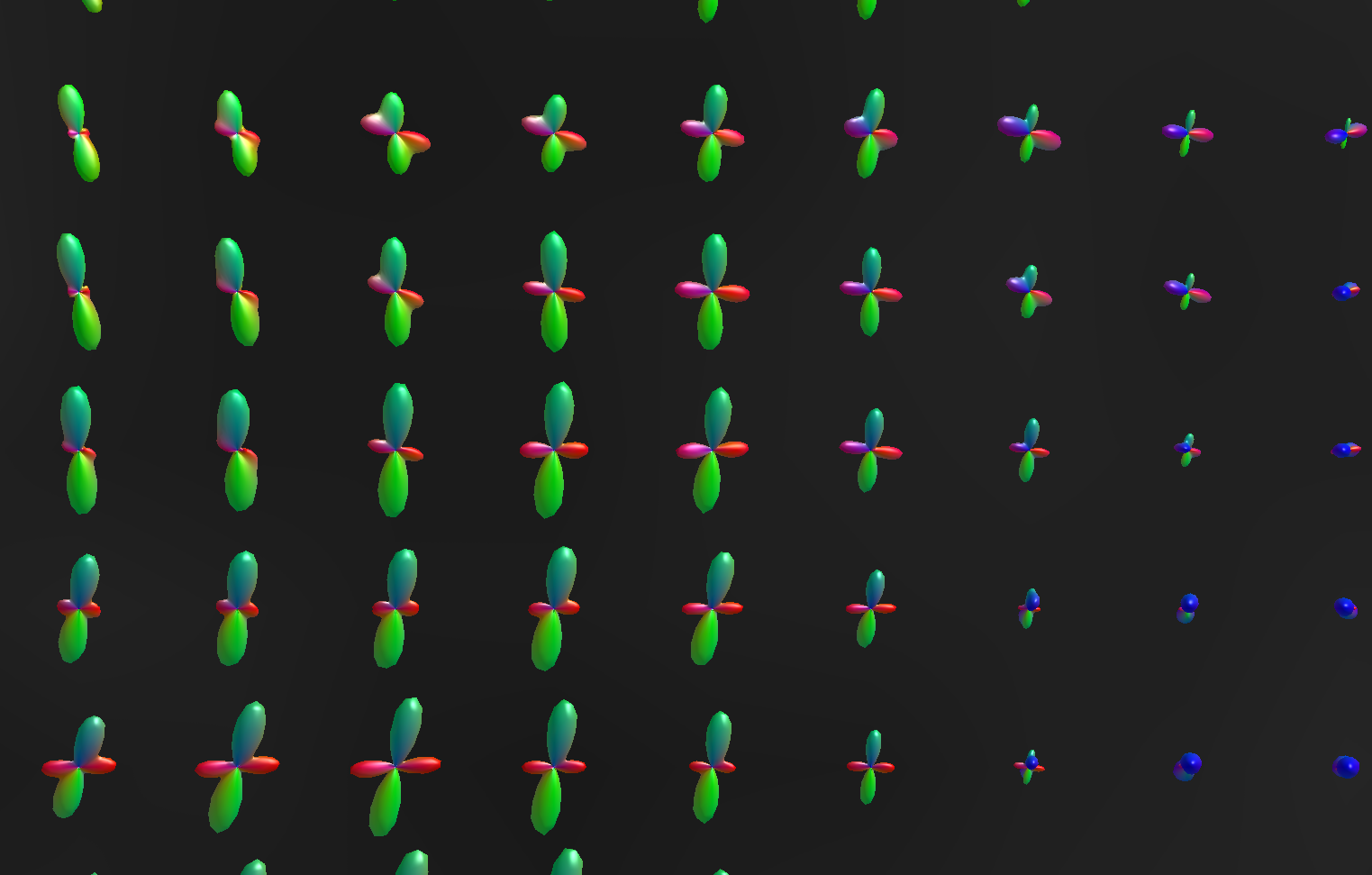}
        \caption{FOD-Swin-Net}
    \end{subfigure}
    \hspace{0.01mm}
    \begin{subfigure}{0.17\textwidth}
        \includegraphics[width=\textwidth, trim=0 0 480 50, clip]{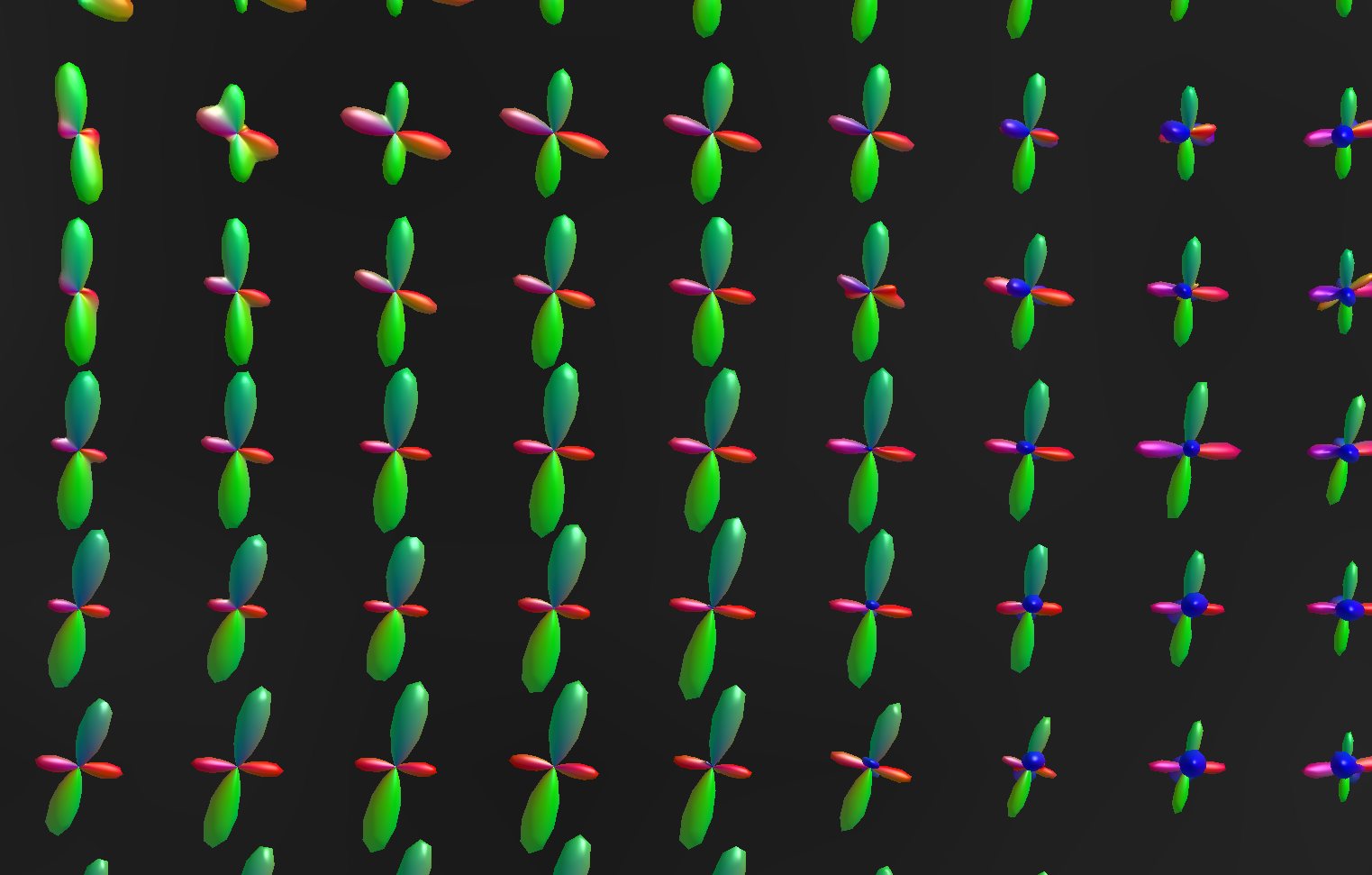}
        \caption{Ground Truth}
    \end{subfigure}
    \caption{Qualitative visualizations in (b-d) showing FOD reconstruction in the region depicted in (a) for SS3T, FOD-Net, and FOD-Swin-Net (this work), with (e) being the multishell-based ground truth. The zoomed-in area corresponds to an intersection between the Corpus Callosum, the Superior Longitudinal Fascicle, and the Arcuate Fascicle - as indicated by the segmentation of the FODs into different fiber bundles using the TractSeg software~\cite{tractseg}}
    \label{fig:qualitative}
\end{figure*}

\begin{figure}[ht]
    \centering
    \begin{subfigure}{0.30\columnwidth}
        \includegraphics[width=\columnwidth]{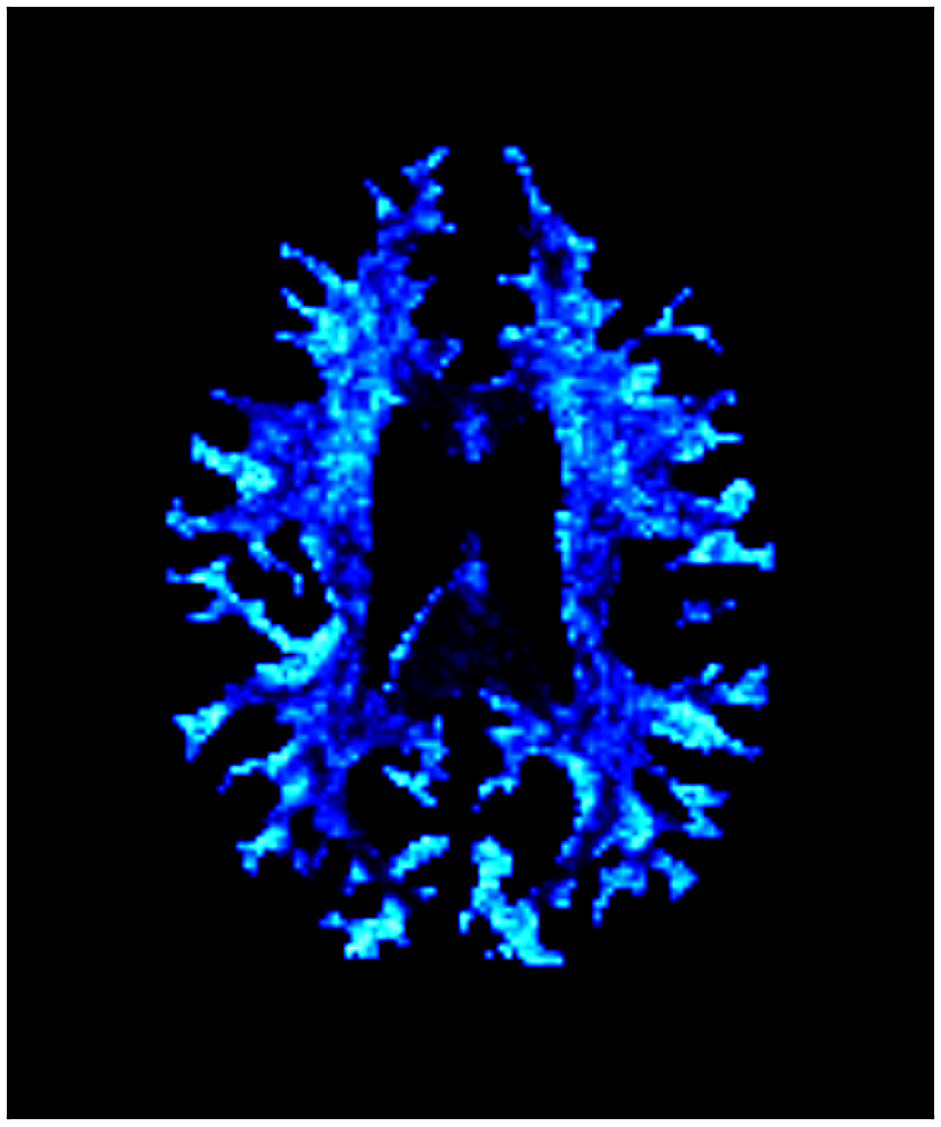}
        \caption{SS3T}
    \end{subfigure}
    \begin{subfigure}{0.30\columnwidth}
        \includegraphics[width=\columnwidth]{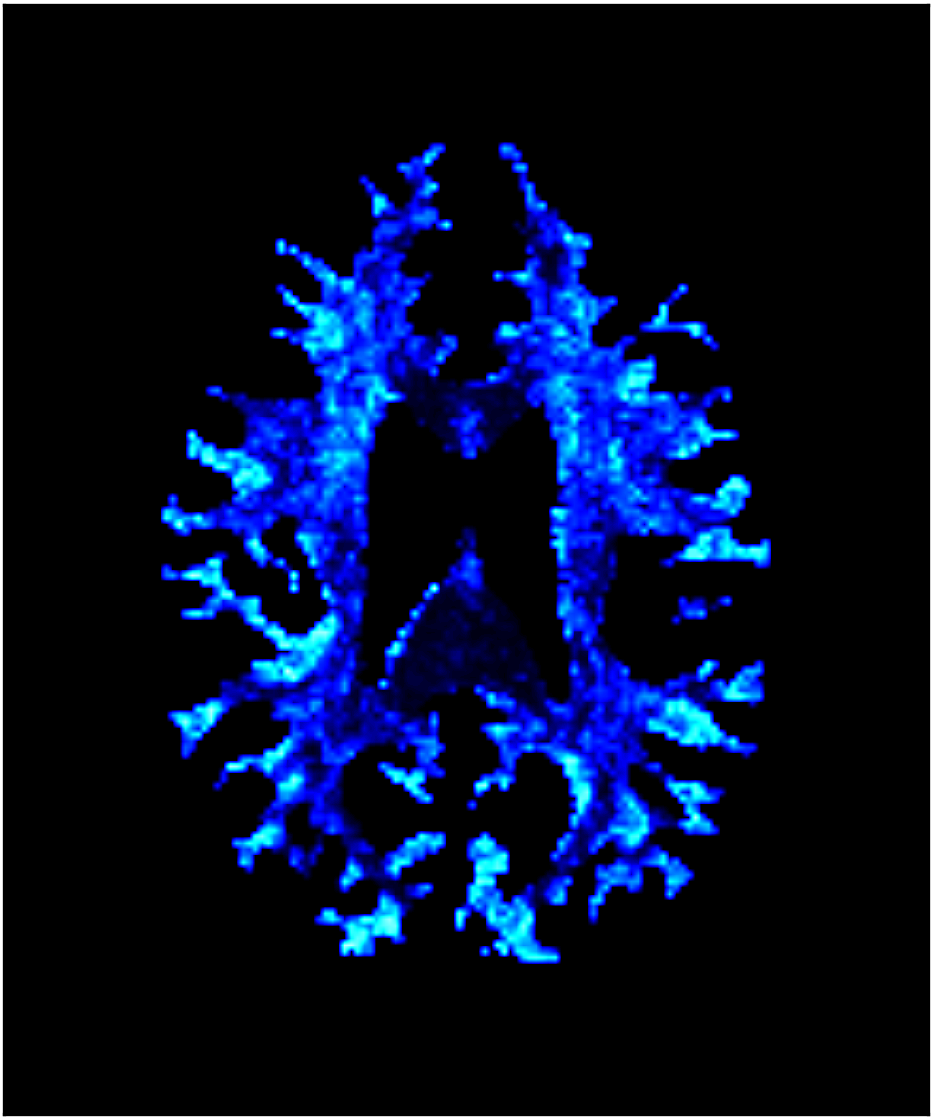}
        \caption{FOD-Net}
    \end{subfigure}
    \begin{subfigure}{0.30\columnwidth}
        \includegraphics[width=\columnwidth]{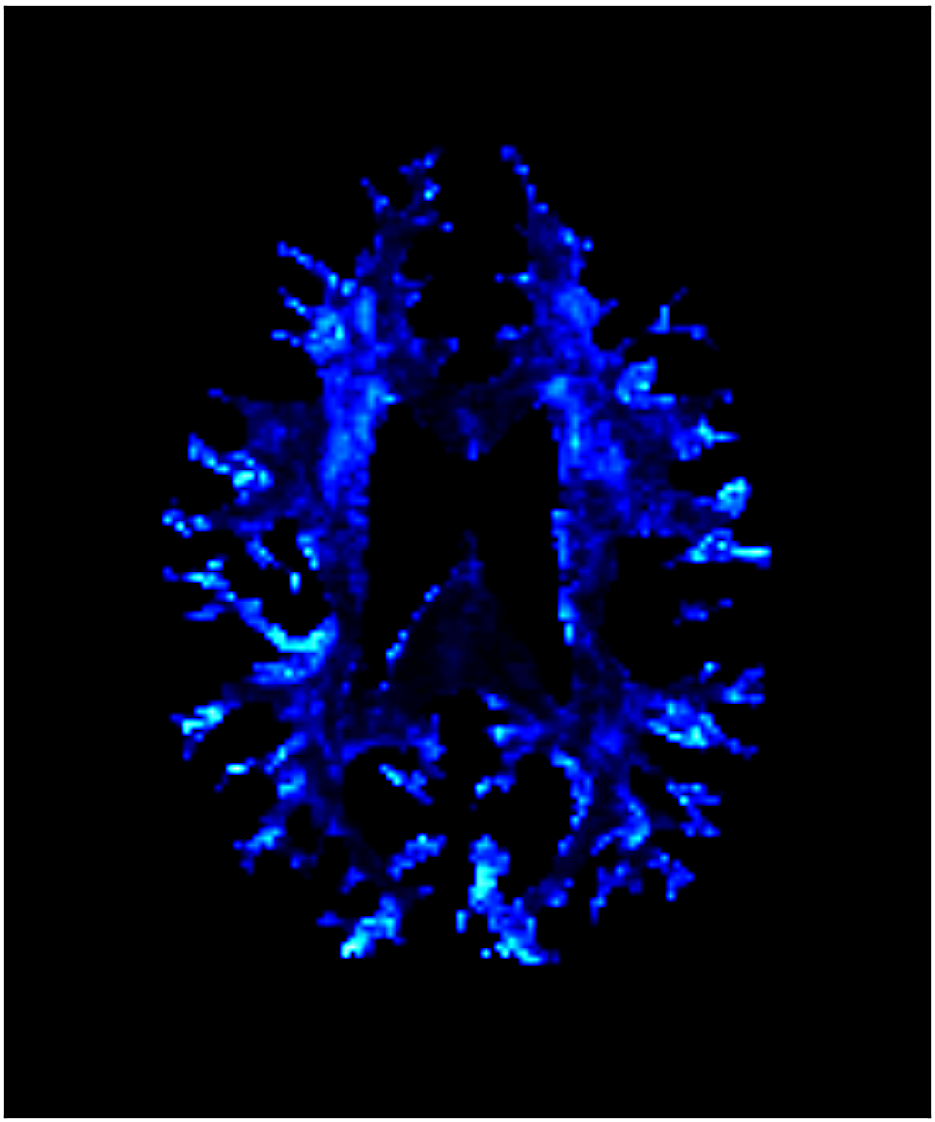}
        \caption{Fod-Swin-Net}
    \end{subfigure}
    \begin{subfigure}{0.062\columnwidth}
        \raisebox{0.6cm}{\includegraphics[width=\columnwidth]{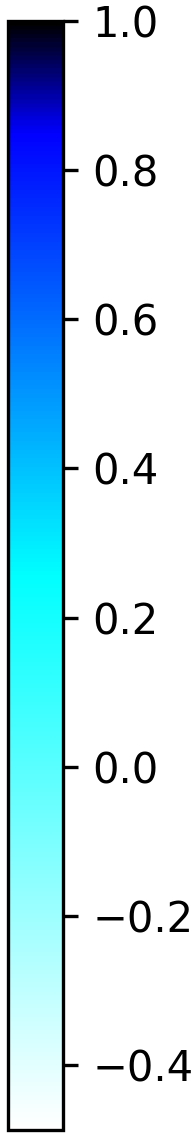}}
    \end{subfigure}

    \caption{ACC heatmap in the same middle WM axial slice from a subject for SS3T (a), FOD-Net (b) and FOD-Swin-Net (c).}
    \label{fig:qualitative_2}
\end{figure}

In addition to the quantitative results, qualitative visualizations of the predicted FOD show better angular super resolution of the fiber zones with less noise when compared to the HARDI ground truth and other methods (Fig.~\ref{fig:qualitative}). Moreover, ACC heatmaps of a central axial slice filtered by WM highlight the lower angular error from reconstructions originating from FOD-Swin-Net (Fig.~\ref{fig:qualitative_2}).

\section{Discussion}

The proposed approach, FOD-Swin-Net, enables the generation of angular super resolution FOD images using single-shell LARDI data. It achieves FOD quality comparable to that of more sophisticated multi-shell HARDI data, while requiring only a fraction of the DWI data.

When evaluating the performance of FOD-Swin-Net quantitatively, we considered the angular accuracy of the predictions (ACC). The FOD-Swin-Net demonstrated significant improvement, achieving the best results in all three tissue types when compared to FOD-Net and SS3T (Table~\ref{metric:table_acc}). The method performed particularly well in the pure WM tissue, but the highest ACC gain was obtained in the WM/CGM tissue (0.18 against SS3T and 0.13 against FOD-Net).


From the ACC distributions (Fig.~\ref{fig:acc_violin_boxplot}), it is clear that FOD-Swin-Net generates more voxels with high ACC values. This is especially evident in the WM/CGM and WM/SGM tissues (b-c), indicating the superior performance of the proposed method on difficult regions affected by partial volume. On these same regions FOD-Net and SS3T distributions almost do not present a peak, with more voxels in lower ACC values.

Besides achieving better quantitative angular coefficients, our method presents visually better spherical harmonic reconstructions (Fig.~\ref{fig:qualitative}). It is visible that FOD-Swin-Net results are consistently similar to the ground truth, while SS3T presents highly discordant FODs (e.g., lower-right corner). Also, FOD-Net results in this region seem like an attenuation of the SS3T reconstruction, keeping its overall shapes and directions but reducing the peaks. Also, the ACC heatmap (Fig.~\ref{fig:qualitative_2}) shows values closer to an optimal maximum correlation in the central regions of the WM, especially in FOD-Swin-Net. Some areas near the borders of the tissue tend to present worse results in SS3T and FOD-Net (lighter regions), which may be due to voxel contamination by other tissue types. However, FOD-Swin-Net is able to greatly attenuate these errors.


We hypothesize the employment of a modern transformer combined with a patch-based encoder-decoder architecture, a different approach than those of voxel-based related work, was paramount in achieving the reported success. Not only the Transformer's state-of-the-art ability to make and analyze global and local representations played an important role in our superior performance, but our 3D patch-wise approach, working with a large spatial neighborhood, lead to much faster sliding window predictions, with FOD-Swin-Net being 60 times faster.


Obtaining higher angular resolution from LARDI data using DL models has the potential to accelerate diffusion acquisitions in the future. However, it is not without limitations. As in FOD-Net, one of the main limitations of this work is requiring coefficients from the SS3T method as starting point.

\section{Conclusion}

We have proposed and evaluated a different approach to FOD angular super resolution, which is able to achieve state-of-the-art performance in angular accuracy and computational speed. FOD-Swin-Net brings single-shell derived FOD representations to be comparable with those generated from 288 directions multi-shell acquisitions. Future work will explore the concept of fixels, taking into consideration the number of crossing fibers and their angular correctness in our analysis. Also, we will include more data sources in future studies.

\section{Compliance with ethical standards}
\label{sec:ethics}

This research study was conducted retrospectively using human subject data made available in open access by the HCP~\cite{van2013wu}. Ethical approval was not required as confirmed by the license attached with the open access data.

\section{Acknowledgments}
\label{sec:acknowledgments}

This study was financed in part by the National Council of Scientific and Technological Development (CNPq). D. Carmo is supported by the São Paulo Research Foundation (FAPESP) grant \#2019/21964-4. Public data used in this work were provided by the Human Connectome Project, WU-Minn Consortium (Principal Investigators: David Van Essen and Kamil Ugurbil; 1U54MH091657) funded by the 16 NIH Institutes and Centers that support the NIH Blueprint for Neuroscience Research; and by the McDonnell Center for Systems Neuroscience at Washington University.

\bibliographystyle{IEEEbib}
\bibliography{refs}

\end{document}